# Corporate sponsorship of computer science conferences: trends, structural insights, and a novel approach to ranking conferences

Yongzhen Wang
yongzhenwang@dlut.edu.cn
School of Public Administration and Policy, Dalian University of Technology, Dalian 116024, China

**Abstract:** Corporate sponsorship is increasingly prevalent at computer science conferences. However, a quantitative understanding of this phenomenon has yet to be established, let alone insights into the interplay between academic conferences and sponsoring corporations, or how to leverage it. To fill these gaps, this study first explores the landscape of corporate sponsorship across a wide range of high-profile computer science conferences, shedding light on its evolution over a 25-year period from 2000 to 2024. The complex and expansive relationships between these conferences and their corporate sponsors are then systematically organized into a network for structural analysis and conference evaluation. Specifically, after modularity optimization, the network's topological properties are analyzed to identify key conferences and corporations that shape the overall structure, connectivity, and functionality. More importantly, this study makes the first attempt to employ a conference–corporation sponsorship network, along with a network-based ranking algorithm, to evaluate computer science conferences, introducing a new perspective on assessing their quality or reputation from the standpoint of corporate sponsorship. The proposed evaluation approach is benchmarked against three popular ranking systems, demonstrating not only its practical usefulness but also its unique ability to highlight the disparity in the attention that academia and industry direct to different fields of computer science. This paper has significant implications for scholarly communication in computer science, particularly as industry has become the primary consumer of academic research in the discipline.

# INTRODUCTION

Academic conferences are integral to the global research landscape, serving as essential platforms for advancing knowledge, fostering collaborations, shaping scientific communities, and driving intellectual progress (de Leon & McQuillin, 2020; Hauss, 2021; Sá et al., 2019). Particularly in computer science, a fast-moving and ever-changing discipline, conferences are often preferred over journals as the primary venue for publication (Freyne et al., 2010; Kim, 2019; Vrettas & Sanderson, 2015). There are numerous prestigious computer science conferences held every year, drawing the attention of individuals and organizations from both academia and industry. Among them, corporations are crucial participants at today's computer science conferences, playing a variety of roles such as business promoters, talent recruiters, innovation exhibitors, and, notably, financial or technical sponsors.

Corporate sponsorship has become prevalent at computer science conferences in recent years. Just like major sporting events, high-profile computer science conferences attract corporate sponsors by offering a unique blend of visibility, networking opportunities, and, most importantly, strategic alignment with their business goals. These sponsorship relationships typically arise when conferences focus on topics that align with the interests of sponsoring corporations. This linkage mechanism motivates an intriguing yet largely unexplored challenge: structuring the complex and expansive relationships between computer science conferences and their corporate sponsors into a network—one that not only clarifies their interplay but also lays the groundwork for addressing a broad spectrum of open problems. For instance, an appealing research direction is to leverage such an innovative conference–corporation sponsorship network to stratify and rank computer science conferences, introducing a new perspective on assessing their quality or reputation from the standpoint of corporate sponsorship.

## Corporate sponsorship of academic conferences

It is widely acknowledged that corporate sponsorship can influence the scientific agenda (Fabbri et al., 2018; Legg et al., 2021). In the context of academic conferences, previous studies—primarily in medicine and public health—have examined its negative effects. For example, Jayasinghe (2021) pointed out that sponsorship of medical conferences by the pharmaceutical industry led to many ethical issues, while Srinivasan (2023) argued that pharmaceutical companies should not be allowed to continue funding doctors' participation in such events. Similarly, Wood et al. (2020) questioned whether current protocols on food industry sponsorship of academic conferences are sufficient to protect public health interests from corporate influence, while Gunnarsson et al. (2023) recommended implementing stricter regulations and safeguards to prevent hidden industry influence on public health conferences and their speakers. In general, these prior studies are either qualitative or centered on case studies of one or more specific corporate sponsors. To date, little research has been conducted on the broader landscape of corporate sponsorship in academic conferences—whether across all disciplines or within a particular one—let alone on the implications and outcomes of industry involvement in these conferences.

## Evaluating computer science conferences

Assessing the quality or reputation of academic conferences has long been a significant challenge in computing disciplines. Although existing bibliometric methods provide a workable solution—similar to their application in journal evaluation—they fail to capture the intrinsic characteristics of computer science conferences (Fortiş & Fortiş, 2024; Meho, 2019). Over the past decades, substantial efforts have been made to develop a diverse array of metrics and indicators to address this issue. A prime example is the work of Martins et al. (2010), who proposed a set of new bibliometric indicators to evaluate computer science conferences. Their approach overcomes the limitations of conventional bibliometric indicators (e.g., h-index) by incorporating critical factors into the evaluation, including longevity, popularity, prestige, and periodicity. On the other hand, Almendra et al. (2015), as well as Effendy and Yap (2016), advocated for the integration of conventional bibliometric indicators with appropriate techniques (e.g., machine learning) to ensure that evaluation results are more effectively tailored to the specific context of computer science conferences.

In addition to the two research strands mentioned above, another promising alternative for evaluating computer science conferences is to leverage characteristics such as program committee composition, submission and acceptance statistics, and peer reputation, rather than relying solely on bibliometric analysis (Küngas et al., 2013; Loizides & Koutsakis, 2017; Martins et al., 2009; Vahdati et al., 2021). These characteristics can serve as useful proxies for assessing the quality or reputation of computer science conferences from a perspective that goes beyond bibliometrics. Consider submission and acceptance statistics as an example: high-profile academic conferences consistently receive a large number of submissions while maintaining relatively low acceptance rates (Küngas et al., 2013; Martins et al., 2009). Likewise, it is plausible that such conferences attract more prominent and influential corporate sponsors on a larger scale. To the best of our knowledge, however, no studies have examined the evaluation of computer science conferences from the standpoint of corporate sponsorship.

### Contributions of this study

This study first explores the landscape of corporate sponsorship across a wide range of well-known computer science conferences, shedding light on its evolution over the past 25 years (2000–2024). This long-term trajectory provides valuable insights into the influence of corporate sponsors on contemporary scholarly communication in computer science. Second, a novel conference–corporation sponsorship network is constructed by systematically organizing the complex and expansive relationships between computer science conferences and their corporate sponsors, in order to better understand the interplay between them. Specifically, after modularity optimization, the network's topological properties are analyzed to identify key conferences and corporations that shape the overall structure, connectivity, and functionality. Third, this study makes the first attempt to employ a conference–corporation sponsorship network, along with a network-based ranking algorithm, to evaluate computer science conferences. The rationale behind this approach is that each sponsorship relationship can be viewed as a form of public, explicit endorsement by the sponsoring corporation, thereby enabling an assessment of a conference's quality or reputation from an industrial perspective. To verify the effectiveness of the proposed evaluation approach, it is benchmarked against three widely recognized ranking systems, demonstrating not only its practical usefulness but also its unique ability to highlight the disparity in the attention that academia and industry direct to different fields of computer science.

## MATERIALS AND METHODS

### Data collection

This study examines corporate sponsorship across a diverse selection of prestigious computer science conferences from 2000 to 2024 (refer to Table S1, in the Supplementary Information). All selected conferences are regarded either as flagship events or as centers of an ecosystem, each of which is ranked A∗ or A (i.e., exceptional or excellent) and is categorized into one or several research fields (e.g., "Artificial intelligence"; see Figure A1, in the Appendix) by the ICORE Conference Portal[1], a collaborative effort between different computer science societies from around the world, primarily focused on assessing the quality or reputation of academic conferences in computing disciplines. The corporate sponsorship of these conferences is manually collected through their web portals or printed proceedings by searching for information on sponsors, supporters, donors, patrons, benefactors, and the like—pure exhibitors are not included. For the sake of clarity in record-keeping, if a sponsor's introductory image or text therein contains both a corporation and its subsidiaries, the sponsorship relationship is attributed only to the parent corporation. If a corporation changes its name, it is treated as a new corporation. If two or more conferences are consistently co-held over the 25-year period (e.g., SIGMOD and PODS), they are considered a single conference, resulting in 172 unique conferences covering 2,670 occurrences (i.e., a conference held in a specific year; e.g., AAAI-2024). Finally, since the purpose of this study is to uncover the patterns of industry involvement in academic conferences, sponsorships from non-profit corporations, as well as media and publishing corporations acting as technical sponsors, are excluded. In total, 19,124 sponsorship relationships are documented involving 168 (≈98%) conferences and 3,474 corporations.

## Network modeling

In this study, an undirected, link-weighted bipartite network is constructed based on the sponsorship relationships between computer science conferences and their corporate sponsors, as illustrated in Figure 1a. More specifically, conferences and corporations represent two disjoint sets of nodes in the network, as indicated by the green and blue circles, respectively, while sponsorship relationships serve as links connecting them, as depicted by the gray lines. The stronger the sponsorship relationship—measured by how frequently a corporation sponsors a conference—the greater the weight of the link, and the thicker the gray line.

---

[1] https://portal.core.edu.au/conf-ranks/

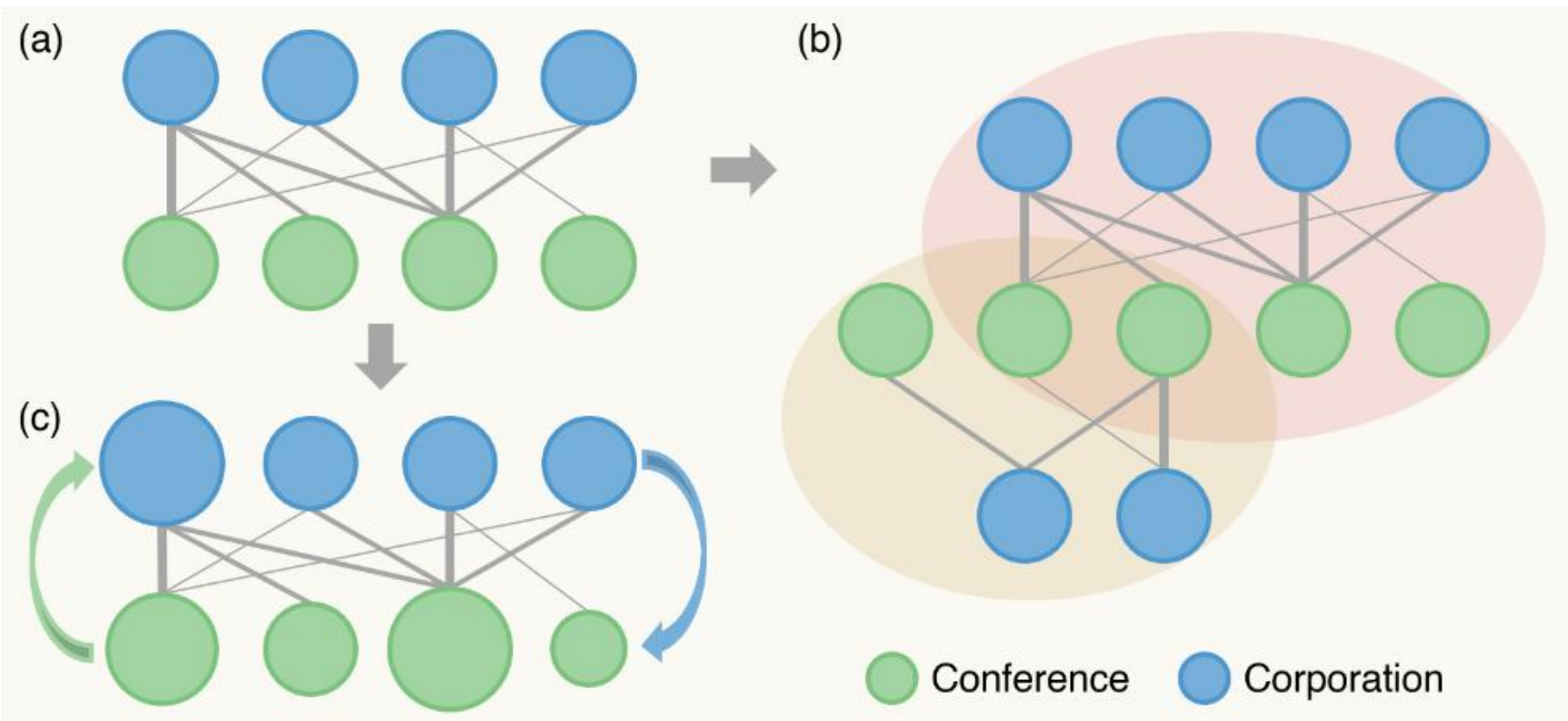

**Figure 1: A toy example illustrating the modeling of a conference–corporation sponsorship network.** (a) An undirected, link-weighted bipartite network; (b) Modularity optimization; (c) Node ranking.

To further investigate the conference–corporation sponsorship network, two well-established network-based analysis algorithms, DIRTLPAwb+ (Beckett, 2016) and BiRank (He et al., 2016), are employed to perform modularity optimization and node ranking, respectively. To be clear, DIRTLPAwb+ is proposed for community detection in weighted bipartite networks, specifically designed to determine the optimal community partition—as indicated by the different colored areas in Figure 1b—while maximizing a metric called weighted modularity, which quantifies the extent to which nodes have stronger connections in their own modules than would be expected if linkages were random. BiRank, a generalized version of PageRank for bipartite networks, generates ranking scores for nodes on both sides simultaneously—as depicted by the different sized circles in Figure 1c—while accounting for the full network topology without information loss, according to the following equations:

$$\boldsymbol{T} = \alpha \boldsymbol{S}_{\mathrm{T}} \boldsymbol{B} + (1 - \alpha) \boldsymbol{T}^{0}$$

$$\boldsymbol{B} = \beta \boldsymbol{S}_{\mathrm{B}} \boldsymbol{T} + (1 - \beta) \boldsymbol{B}^{0}$$

where $\boldsymbol{T}$ and $\boldsymbol{B}$ are the ranking scores for conference and corporation nodes, elements in $\boldsymbol{T}^0$ and $\boldsymbol{B}^0$ are set to $1/|\boldsymbol{T}|$ and $1/|\boldsymbol{B}|$ by default, α and β are damping factors whose values range between 0 and 1, and $\boldsymbol{S}_{\mathrm{T}}$ and $\boldsymbol{S}_{\mathrm{B}}$ are the transition matrices. In this study, BiRank is executed multiple times, with the values of α and β varying from 0.6 to 0.95 in intervals of 0.05, and the final ranking scores, referred to hereafter as BiRank scores, are calculated as the average of the results from these multiple runs.

Consequently, we are able to identify the role of each node in the network (i.e., hub, connector, or peripheral) and assess its relative importance (i.e., network centrality). As a side note, DIRTLPAwb+ and BiRank are implemented in R using the "bipartite" and "birankr" packages (Dormann et al., 2008; Yang et al., 2020).

## RESULTS AND INTERPRETATIONS

### Evolving landscape of corporate sponsorship at computer science conferences

Before delving into the conference–corporation sponsorship network, let us first explore the landscape of corporate sponsorship at computer science conferences over the past 25 years (2000–2024). To facilitate analysis, this 25-year period is divided into five equal-length eras and examined from three perspectives: *Density*, *Coverage*, and *Diversity*, as shown in Figure 2. More specifically, *Density* refers to the number of corporate sponsors per sponsored conference, *Coverage* refers to the percentage of conferences receiving corporate sponsorship, and *Diversity* refers to the number of distinct corporate sponsors.

Overall, all three perspectives exhibit an upward trend, despite a moderate decline in the most recent era (2020–2024). The most dramatic increase in *Coverage* takes place between 2005 and 2009, rising from under 50% to over 70%, coinciding with evidence that major tech companies were increasingly recognizing conferences as strategic platforms (Baruffaldi & Poege, 2025). *Diversity* sees its greatest increase between 2015 and 2019, almost doubling in magnitude, indicating a substantial expansion in the pool of corporate sponsors during this span. As for *Density,* the situation is somewhat more complicated: the median remains relatively stable, but the range of values expands considerably—especially between 2015 and 2019—due to a soaring maximum. This finding suggests a highly skewed distribution, where a small number of high-profile conferences capture a growing share of sponsorship. For instance, CVPR-2017 (the 30th IEEE/CVF Conference on Computer Vision and Pattern Recognition) has more than 100 corporate sponsors, the most among all occurrences included in this study. By and large, corporate sponsorship of computer science conferences has grown significantly since 2000. However, is this growth consistent across all fields of computer science?

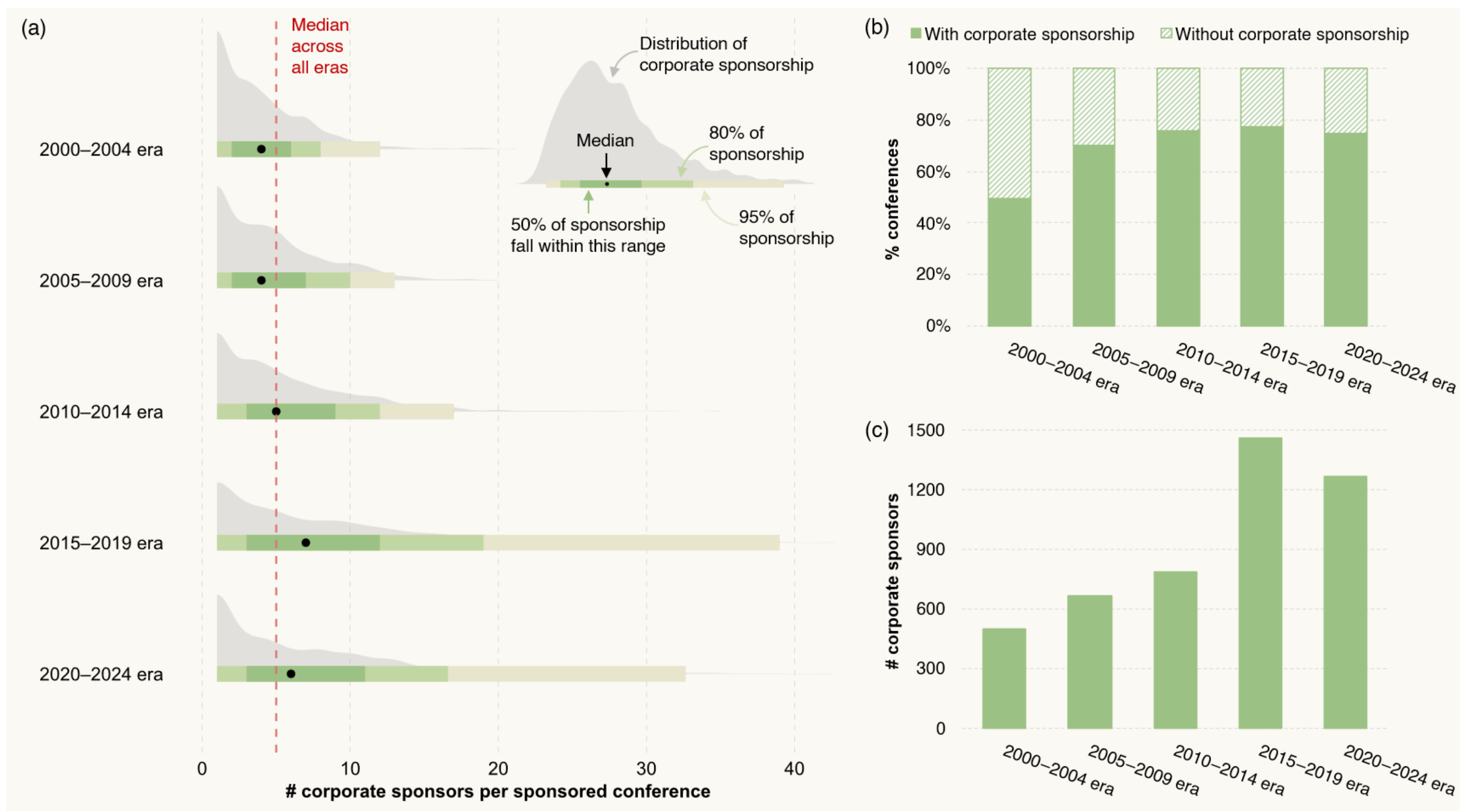

**Figure 2: Trends in corporate sponsorship from three perspectives.** (a) *Density*; (b) *Coverage*; (c) *Diversity*. For ease of viewing, the distributions in (a) are truncated to exclude outliers.

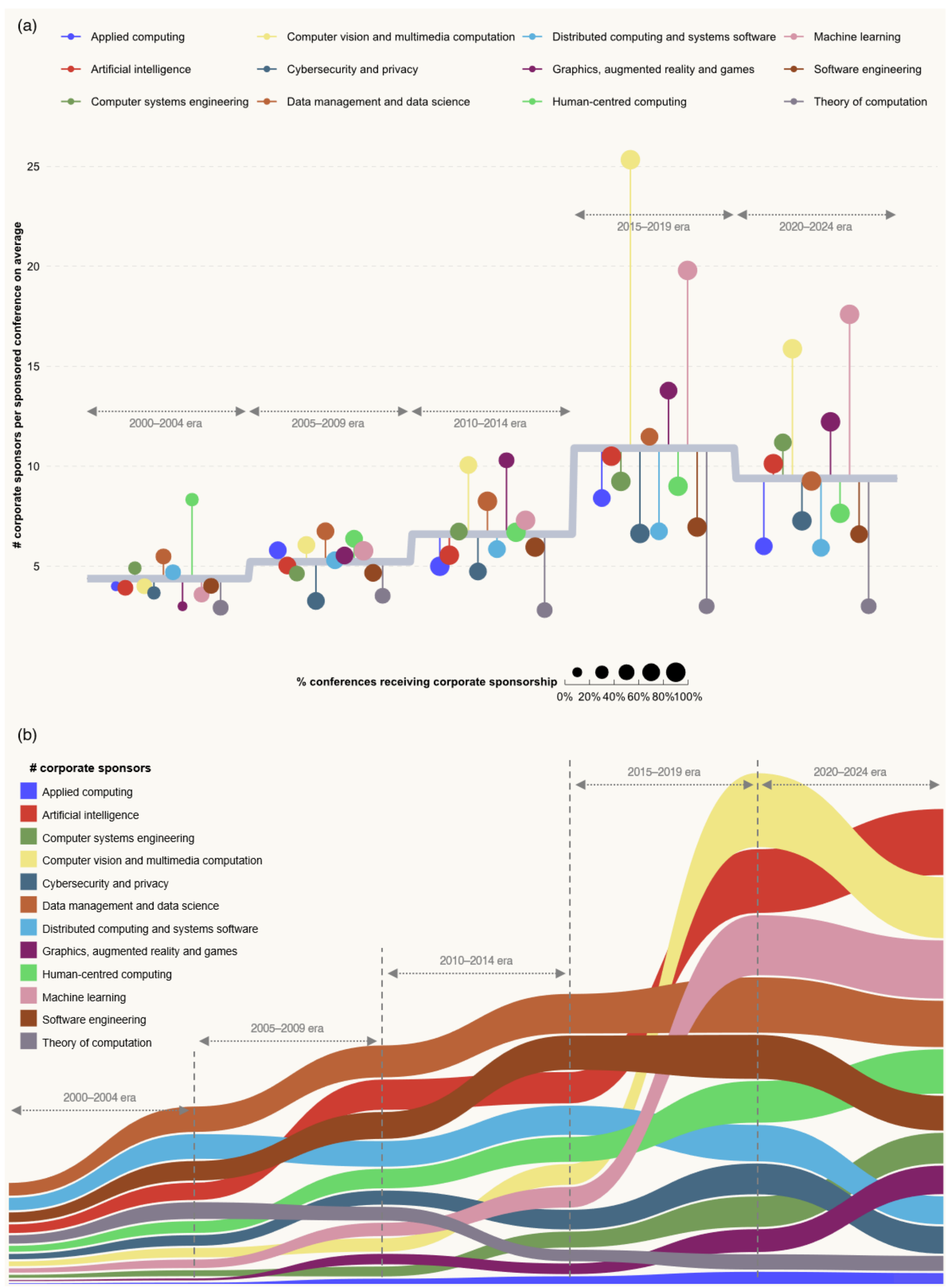

**Figure 3: Field-specific trends in corporate sponsorship from three perspectives.** (a) *Density* and *Coverage*; (b) *Diversity*.

To illustrate the variations across fields, Figure 3 displays the field-specific trends in corporate sponsorship of computer science conferences. Evidently, substantial differences exist, particularly in terms of *Density* and *Diversity*, although differences in *Coverage* tend to diminish over time. More specifically, each of the two fields—"Computer vision and multimedia computation" and "Machine learning"—has witnessed enormous increases in both *Density* and *Diversity* between 2015 and 2024. In contrast, "Theory of computation" shows no significant growth in *Density* from 2000 to 2024, accompanied by a marked decrease in *Diversity*. Additionally, "Applied computing" has maintained the lowest level of *Diversity* over the 25-year period, whereas "Artificial intelligence" has experienced a steady upward trend in *Diversity*, reaching its highest level between 2020 and 2024. One final noteworthy point is that, while "Graphics, augmented reality and games" exhibits a relatively high *Density* between 2010 and 2024, it ranks low in *Diversity*—corporate sponsors in this field have remained fairly constant over the course of 15 years.

The moderate decline in corporate sponsorship of computer science conferences between 2020 and 2024 may be attributed in part to COVID-19. In response to the pandemic, numerous computer science conferences transitioned to an online format between 2020 and 2023 (i.e., the duration of the WHO public health emergency declaration). The shift to an online format brought about a series of changes, such as lower registration fees (Mubin et al., 2021), fewer networking opportunities (Fraser & Mancl, 2024), and possibly reduced visibility for corporate sponsors. These circumstances may have diminished the perceived value of sponsorship, leading in turn to decreased industry involvement—though whether this causal inference is valid requires further investigation.

## Structural characteristics of the conference–corporation sponsorship network

Next, let us explore the conference–corporation sponsorship network in greater detail. As shown in Figure 4, on average, conference nodes have higher degree centrality than corporation nodes, indicating that conferences connect to more corporate sponsors than vice versa. More importantly, the overall degree distribution of nodes approximates a power-law pattern, with an estimated exponent of 2.10, as determined by the maximum likelihood method. Therefore, according to the definition provided by Barabási and Albert (1999), the conference–corporation sponsorship network indeed exhibits scale-free properties. In such networks, the Matthew effect (Merton, 1968) is evident: while most conferences/corporations are connected to only a few others, some function as "super hubs," linking to the majority of corporations/conferences due to their broad-ranging topics/interests. Remarkably, this phenomenon occurs even when considering only high-profile conferences. Furthermore, the inequality in the number of sponsorship relationships among corporations—measured by the Gini coefficient ($G$), where $G$=0 represents perfect equality and $G$=1 represents maximal inequality—is both glaring and greater than that among conferences ($G_{\text{corp.}}$=0.76, $G_{\text{conf.}}$=0.49). To be specific, fewer than 20% of corporations account for 80% of sponsorship relationships—roughly in line with the Pareto principle (Sanders, 1987)—whereas nearly 40% of conferences account for 80% as well. Now, here comes a question: how do conferences and their corporate sponsors engage with each other in a disproportionate manner?

To achieve a deeper understanding of the structure of the conference–corporation sponsorship network, the standardized participation and connection values—denoted as $z$ and $c$—are calculated for each node to describe its role in the network. This follows the methodology proposed by Olesen et al. (2007), which was used to distinguish the roles of species in pollinator networks. More specifically, $z$ refers to the within-module degree, and $c$ refers to the among-module connectivity; both are quantified based on link weights and module partitions. For greater clarity, the calculations of $z_o$ and $c_o$ for a given node $o$ proceed as follows:

$$z_o = \frac{w_{os} - \mu_s}{\sigma_s}$$

where $w_{os}$ is the sum of weights for links from node $o$ to the other nodes in its own module $s$, and $\mu_s$ and $\sigma_s$ are the mean and standard deviation of $w_{*s}$ for each node in module $s$.

$$c_o = 1 - \sum_{t=1}^{m} \left(\frac{w_{ot}}{w_{o*}}\right)^2$$

where $m$ is the number of modules in the network, $w_{n*}$ is the sum of weights for links incident to node $o$, and $w_{ot}$ is the sum of weights for links from node $o$ to nodes in module $t$ (including its own module).

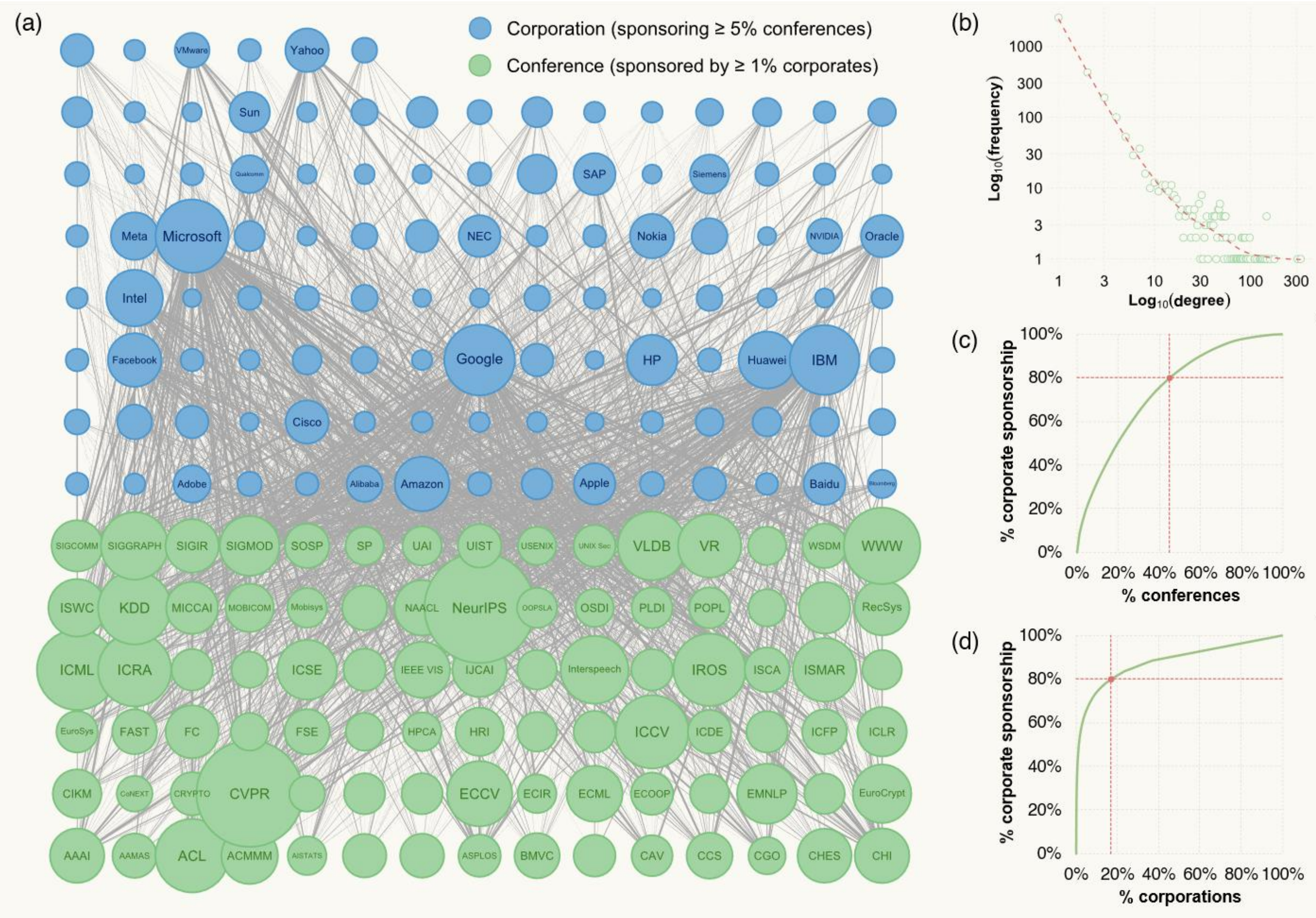

**Figure 4: Network structural characteristics.** (a) Simplified structure of the conference–corporation sponsorship network. Here, the size of each node is proportional to its degree, the width of each link is proportional to its weight, and labeled nodes are those with at least 100 sponsorship relationships. (b) Degree distribution of nodes in a log-log plot. (c)–(d) Lorenz curves showing inequality in the number of sponsorship relationships among conferences and among corporate sponsors, respectively.

Basically, the role of each node in the network is determined by its topological properties—specifically, its position relative to other nodes in its own module, as measured by $z$, and how well it connects to nodes in other modules, as measured by $c$ (Guimera & Nunes Amaral, 2005). As illustrated in Figure 5, $z$ and $c$ together define each node's role via the $z$-$c$ parameter space. All nodes are classified as follows: peripherals (low $z$, low $c$), which have only a few links, almost always to nodes in their own modules; connectors (low $z$, high $c$), which link several modules; module hubs (high $z$, low $c$), which are densely connected nodes that link to many others in their own modules; and network hubs (high $z$, high $c$), which function as both connectors and module hubs.

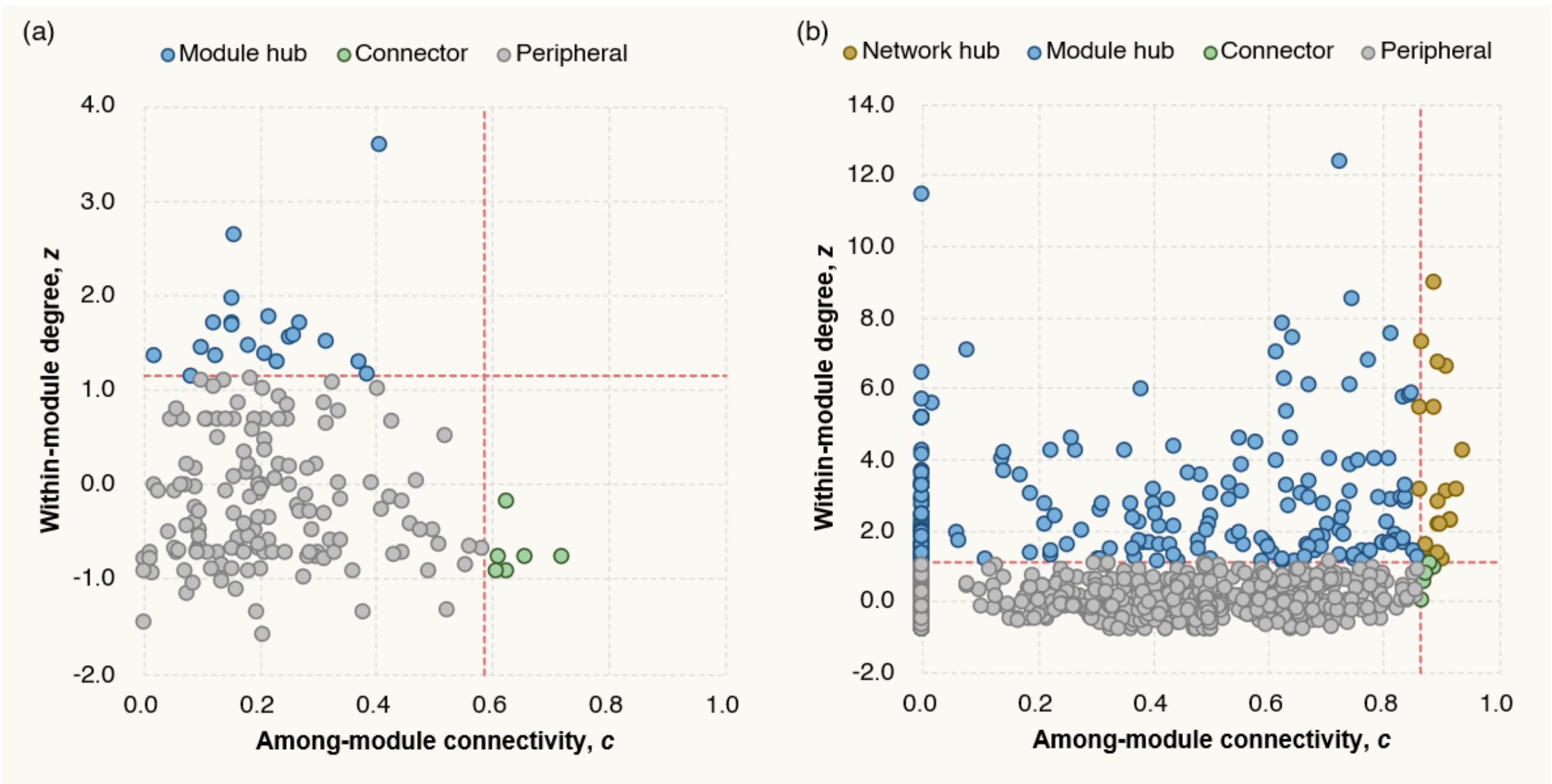

**Figure 5: Node role distributions.** (a) Conferences; (b) Corporate sponsors. Red dashed lines indicate critical values derived from 100 null models.

For conference nodes, 84.5% are peripherals, 11.9% are module hubs, and 3.6% are connectors; however, none are classified as network hubs. As for corporation nodes, 92.2% are peripherals, 7.2% are module hubs, 0.5% are network hubs, and 0.1% are connectors (difference between the two node types: $\chi^2$=128.5, $df$=3, $p$<0.001). Collectively, the network's backbone is maintained by certain conferences and corporations that serve as network hubs or module hubs (see Figure 4a). They are also the fundamental force shaping public perception of corporate sponsorship at computer science conferences (Baruffaldi & Poege, 2025). Meanwhile, information sharing and coordination within the network are primarily driven by an even smaller group of conferences and corporations that serve as connectors. They are typically cross-cutting, either in terms of focal topics for conferences or business interests for corporations. For example, COLT (Annual Conference on Learning Theory) bridges its core subject with related topics such as statistics, optimization, and game theory, connecting diverse research communities; ABB (ASEA Brown Boveri Ltd.) connects multiple industrial sectors such as electrification and automation while maintaining active collaborations with academic institutions. Last but not least, the remaining entities, though acting as peripherals, are by no means trivial. On the contrary, they make up the majority and are the very reason the network exists and operates. In fact, they contribute significantly to expanding the boundaries of academic sponsorship within the industrial sphere, ultimately enriching the global research landscape. In summary, the conference–corporation sponsorship network centers on a handful of key conferences and corporations, surrounded by a vast periphery, with the interactions between these central and peripheral elements forming the foundation of the network's infrastructure.

## Ranking computer science conferences based on corporate sponsorship

Finally, let us examine whether combining the conference–corporation sponsorship network with the BiRank algorithm, the proposed evaluation approach, can effectively rank computer science conferences.

Figure 6 presents the rankings of conferences and their corresponding fields based on BiRank scores. It can be observed that two conferences (i.e., CVPR and NeurIPS) have much higher BiRank scores than the others, and as the conference rankings drop, the differences in BiRank scores between successive conferences become smaller. This pattern echoes the network's degree distribution (see Figure 4b), confirming the existence of a hierarchy among conferences regarding quality or reputation from the standpoint of corporate sponsorship, with a few emerging as clear leaders. In addition, BiRank scores vary significantly across fields. For example, "Graphics, augmented reality and games," "Computer vision and multimedia computation," and "Machine learning" rank among the top three, each demonstrating an obvious advantage over the remaining fields. These fields are precisely those with consistently the highest *Density* levels since 2005 (see Figure 3a), suggesting a strong alignment between corporate sponsorship volume and inferred conference quality. In contrast, "Theory of computation" ranks last, with an average BiRank score considerably lower than that of any other field. Notwithstanding the strong discriminatory power, the extent to which BiRank scores are reliable is unclear.

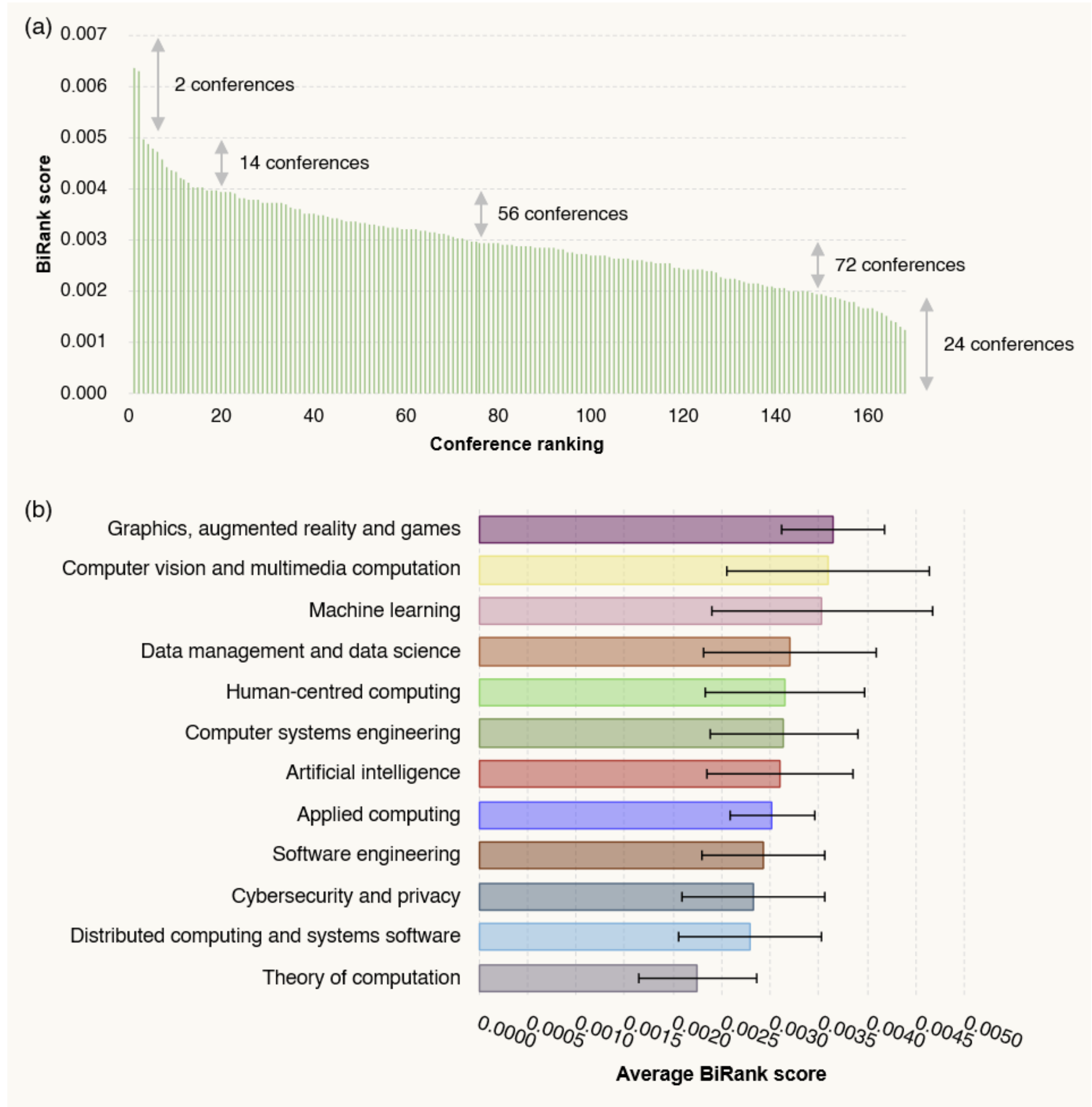

**Figure 6: BiRank scores**. (a) Conferences; (b) Fields. Error bars in (b) indicate standard deviations.

To validate the accuracy and robustness of the proposed evaluation approach, three widely recognized computer science conference ranking systems are selected as benchmarks: the ICORE Conference Portal, Research.com[2], and AMiner[3].

---

[2] https://research.com/conference-rankings/computer-science

[3] https://www.aminer.cn/ranks/conf

Research.com is an online platform that compiles academic rankings and educational resources, including a ranking list of computer science conferences. AMiner is an online academic search and mining system, with a particular focus on computer science. Unlike the ICORE Conference Portal, which assigns each conference an ordinal rank (i.e., A∗, A, B, or C), Research.com and AMiner evaluate conferences using numerical measures: impact scores and h5-index values, respectively. To clarify, ICORE ranks are determined based on a combination of criteria, including citation rates, paper submission and acceptance rates, and the visibility and research track record of the key individuals hosting the conference and managing its technical program. Impact scores take into account both the number of contributing scientists and the h-index estimated from their published work. The h5-index is a version of the h-index calculated specifically for articles published within the last five complete calendar years. By the way, ICORE ranks, impact scores, and h5-index values all reflect a conference's quality or reputation within the academic sphere, whereas BiRank scores indicate its standing within the industrial sphere.

Figure 7 compares A∗ and A-ranked conferences across fields based on BiRank scores. It is evident that A∗-ranked conferences consistently have higher BiRank scores than A-ranked conferences, whether across all fields or within a particular one. This result suggests that BiRank scores generally correlate well with ICORE ranks, lending initial support to the soundness of the proposed evaluation approach. Among A∗-ranked conferences, the top four fields with the highest BiRank scores are "Computer vision and multimedia computation," "Data management and data science," "Machine learning," and "Graphics, augmented reality and games"—an ordering that differs slightly from that shown in Figure 6. Moreover, the differences in BiRank scores between A∗ and A-ranked conferences vary greatly across fields. For example, the largest gap appears in "Computer vision and multimedia computation" ($\Delta$=0.0013), whereas "Theory of computation" shows no significant difference ($\Delta$=0.0001). These variations demonstrate the ability of BiRank scores to capture nuanced, field-specific distinctions in industry sponsorship.

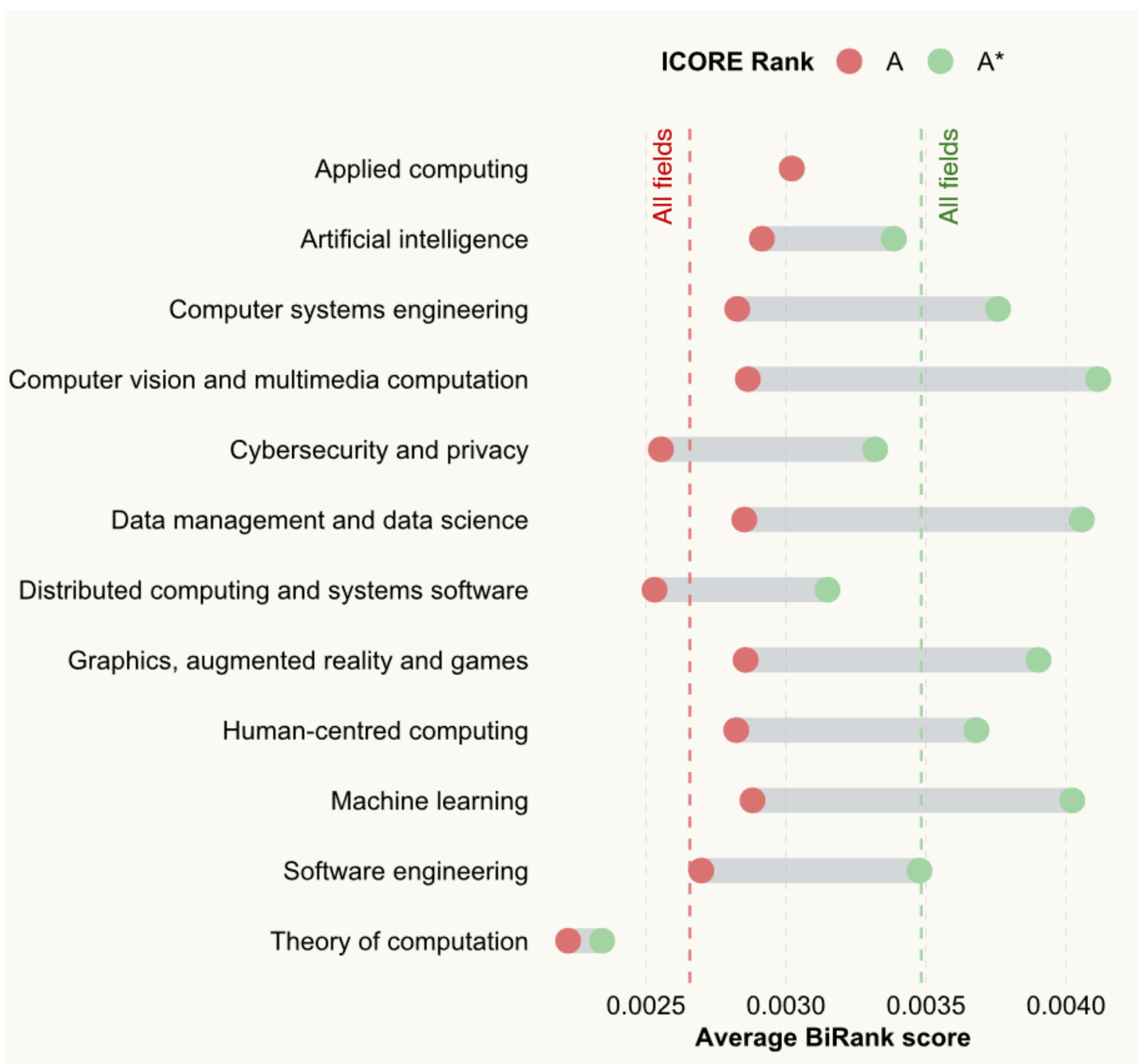

**Figure 7: Comparison of A∗ and A-ranked conferences based on BiRank scores across fields.** No conferences in "Applied computing" are ranked A∗ by the ICORE Conference Portal.

Table 1 lists both the overall and field-specific correlations of BiRank scores with impact scores and h5-index values. Note that some conferences included in this study are not evaluated by Research.com or AMiner, and therefore lack reported impact scores or h5-index values. On the whole, BiRank scores reveal moderate correlations with impact scores (Spearman's $\rho$=0.60) and h5-index values ($\rho$=0.55), as interpreted using the guidelines provided by Schober et al. (2018). Out of the 12 fields of interest, six (50%) exhibit strong correlations between BiRank scores and impact scores ($|\rho|\geq 0.70$), three (25%) show moderate correlations ($0.40\leq|\rho|<0.70$), and the remaining three (25%) display weak correlations ($0.10\leq|\rho|<0.40$). Likewise, for the correlations between BiRank scores and h5-index values, six fields (50%) exhibit strong correlations, three (25%) moderate, and three (25%) weak. Altogether, BiRank scores correlate reasonably well with both impact scores and h5-index values. Nevertheless, they consistently show poor correlations in two fields: “Software engineering” and “Theory of computation,” indicating that these two fields receive considerably less industry attention, resulting in weaker alignment between corporate sponsorship volume and academic prestige.

**Table 1: Correlations between BiRank scores and two established metrics for evaluating computer science conferences.**

| | Impact score | | H5-index | |
|---|---|---|---|---|
| **Field** | **Spearman's ρ** | ***N*** | **Spearman's ρ** | ***N*** |
| Applied computing | 0.89* | 5 | 0.50 | 3 |
| Artificial intelligence | 0.80*** | 23 | 0.78*** | 17 |
| Computer systems engineering | 0.77* | 7 | 0.85** | 9 |
| Computer vision and multimedia computation | 0.45 | 12 | 0.70* | 10 |
| Cybersecurity and privacy | 0.32 | 16 | 0.45 | 11 |
| Data management and data science | 0.61** | 24 | 0.72*** | 22 |
| Distributed computing and systems software | 0.74*** | 33 | 0.59*** | 30 |
| Graphics, augmented reality and games | 0.80 | 4 | 1.00 | 3 |
| Human-centred computing | 0.65** | 15 | −0.18 | 9 |
| Machine learning | 0.77** | 14 | 0.74** | 12 |
| Software engineering | 0.31 | 29 | 0.26 | 20 |
| Theory of computation | 0.10 | 25 | 0.18 | 16 |
| All | 0.60*** | 160 | 0.55*** | 127 |

* $p<0.05$, ** $p<0.01$, *** $p<0.001$

Taken together, the proposed evaluation approach demonstrates satisfactory alignment with the three widely recognized computer science conference ranking systems. What sets it apart, however, are the novel features it introduces, which complement these existing systems by deriving an alternative indicator from corporate sponsorship.

To highlight the distinctiveness of the proposed approach for evaluating computer science conferences, Figure 8 uses the h5-index as an example to illustrate the differences between BiRank scores and h5-index values across fields. For ease of comparison, both metrics are z-normalized. Obviously, h5-index values exhibit greater inequality across fields than BiRank scores ($G_{\text{BiRank}}$=0.07, $G_{\text{h5-index}}$=0.33), with two fields—“Computer vision and multimedia computation” and “Machine learning”—showing significantly higher average h5-index values than the others. This concentration reflects academia’s disproportionate attention to a select few fields. In contrast, BiRank scores are relatively evenly distributed across fields, except for “Theory of computation,” which stands out due to its notably low score. The most pronounced disparities between the two metrics are observed in “Human-centred computing” and “Graphics, augmented reality and games,” both of which receive greater recognition from industry than from academia. These fields, while valued in academia, appear to be particularly attractive to corporate sponsors. Furthermore, aside from the four (33%) fields—“Computer vision and multimedia computation,” “Cybersecurity and privacy,” “Machine learning,” and “Theory of computation”—the other eight (67%) fields are valued more heavily by industry than by academia. This finding points to a broader diffusion of industry attention relative to academia. Similarly, a comparison between BiRank scores and impact scores is provided in Figure A2, in the Appendix. Both comparisons can enhance our understanding of the gap between academia and industry in computer science, particularly with respect to their respective trending fields.

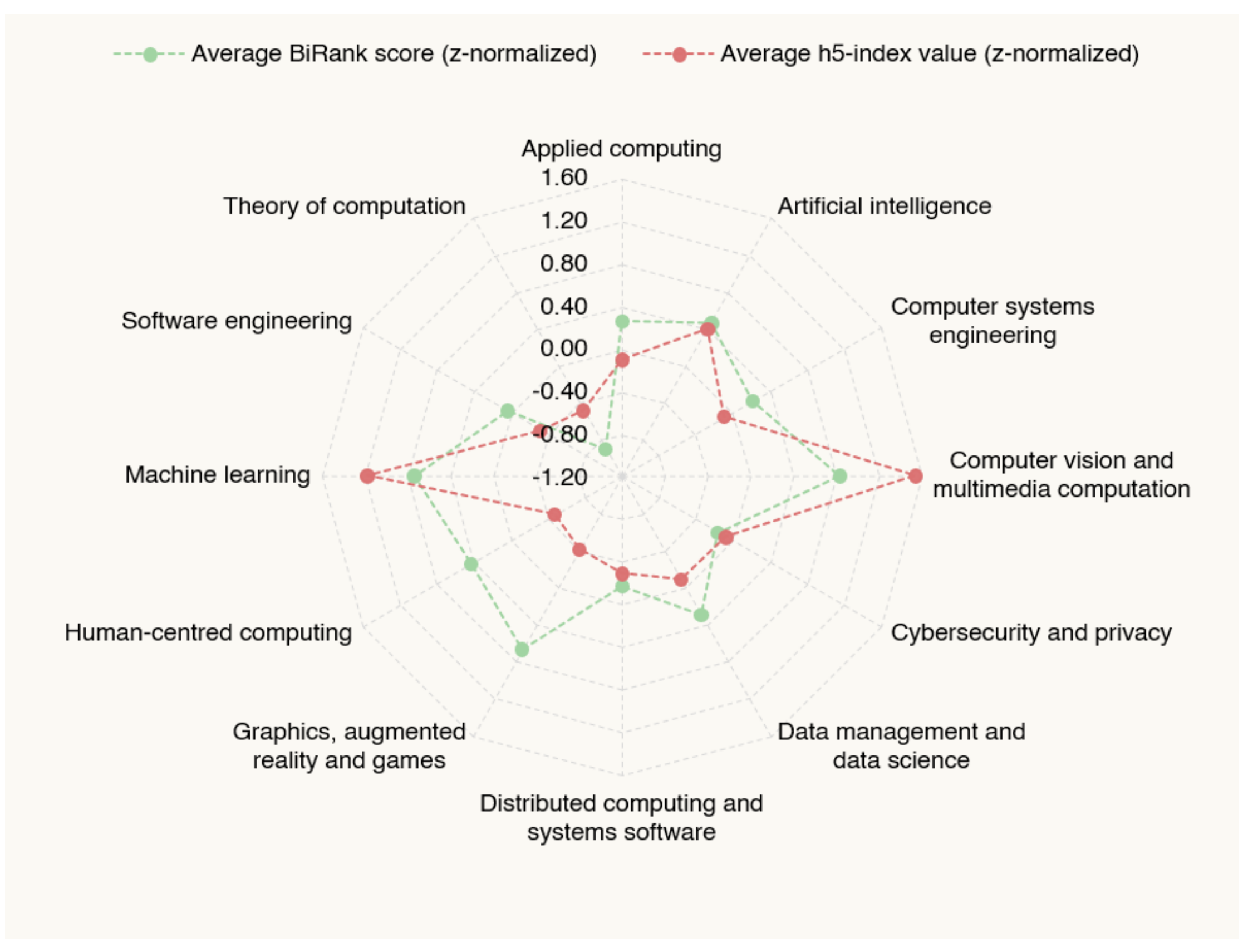

**Figure 8: Average BiRank scores versus h5-index values across fields.**

## CONCLUSIONS AND FUTURE WORK

First of all, this study clearly reveals a remarkable boom in corporate sponsorship of computer science conferences over a 25-year period from 2000 to 2024. Corporate sponsors have evolved from being mere participants to becoming, perhaps, major stakeholders at computer science conferences (Baruffaldi & Poege, 2025). One possible explanation is that industry has become the primary consumer of academic research in computer science. Recently, this trend has raised concerns about the potential risks associated with industry control over computer science conferences, an issue that deserves further exploration. In particular, when industry exerts strong influence on conference funding and organization, it may inadvertently shape research agendas and limit the scope of scholarly inquiry (Jazwinska, 2022). Moreover, it is worth noting that conferences in different fields of computer science vary greatly in their appeal to corporate sponsors, a finding that underscores the uneven distribution of industry attention across the discipline. For example, the three fields—"Artificial intelligence," "Computer vision and multimedia computation," and "Machine learning"—are increasingly favored by corporate sponsors, whereas "Theory of computing" has long been less favorably regarded.

Second, this study elucidates the interplay between computer science conferences and their corporate sponsors by structuring the complex and expansive relationships between them into a network, which is shown to exhibit scale-free properties. In this network, sponsorship relationships are highly concentrated, both among conferences and among corporations. Specifically, a small subset of conferences and corporations dominate the evolving landscape of corporate sponsorship. From a topological standpoint, considering their within-module degree and among-module connectivity (measured via modularity optimization), these entities function as central hubs or modular connectors, exerting disproportionate influence by facilitating interactions or bridging otherwise isolated groups. In other words, they orchestrate the flow of resources and knowledge across the network. Undoubtedly, these entities are of significant importance and merit deeper investigation to fully uncover their broader implications, especially the underlying mechanisms that drive their behavior, as well as the downstream consequences of their dominance for the diversity and openness of scholarly communication.

Third, this study is the first to employ a conference–corporation sponsorship network, along with a network-based ranking algorithm, to evaluate computer science conferences. The empirical results demonstrate that the proposed evaluation approach aligns satisfactorily with three widely recognized ranking systems: the ICORE Conference Portal, Research.com, and AMiner, all of which emphasize quality or reputation from an academic perspective. More importantly, the proposed approach introduces a new pathway for evaluating computer science conferences from an industrial perspective—specifically, through the lens of corporate sponsorship—and serves as a valuable complement to existing ranking systems. By doing so, it reflects the strategic interests and priorities of industry, which are often overlooked in purely academic assessments. Comparing it with these systems also helps illuminate the varying levels of attention that academia and industry direct to different fields of computer science. This comparative analysis reveals where industry attention is concentrated and where it remains limited, providing a granular understanding of the academia–industry gap. In the future, more sophisticated network-based ranking algorithms could be adopted to enhance the proposed evaluation approach, such as incorporating corporate characteristics as prior knowledge to better capture the heterogeneity and dynamics of industry participation.

As an additional point, another avenue for research could be to broaden the scope of corporate sponsorship to include lesser-known computer science conferences, even though these conferences rarely receive such support. Such an extension would provide a more complete picture of how industry attention is distributed across the entire conference landscape, beyond the highly visible flagship events. Note that collecting data on corporate sponsorship of computer science conferences can be both time-consuming and challenging, especially for periods prior to 2005, when such data becomes extremely difficult to obtain. Because of this, the findings presented in this paper should be interpreted with caution. Nevertheless, this study offers a robust foundation for future research by establishing a systematic framework for analyzing conference–corporation sponsorship networks.

# APPENDIX

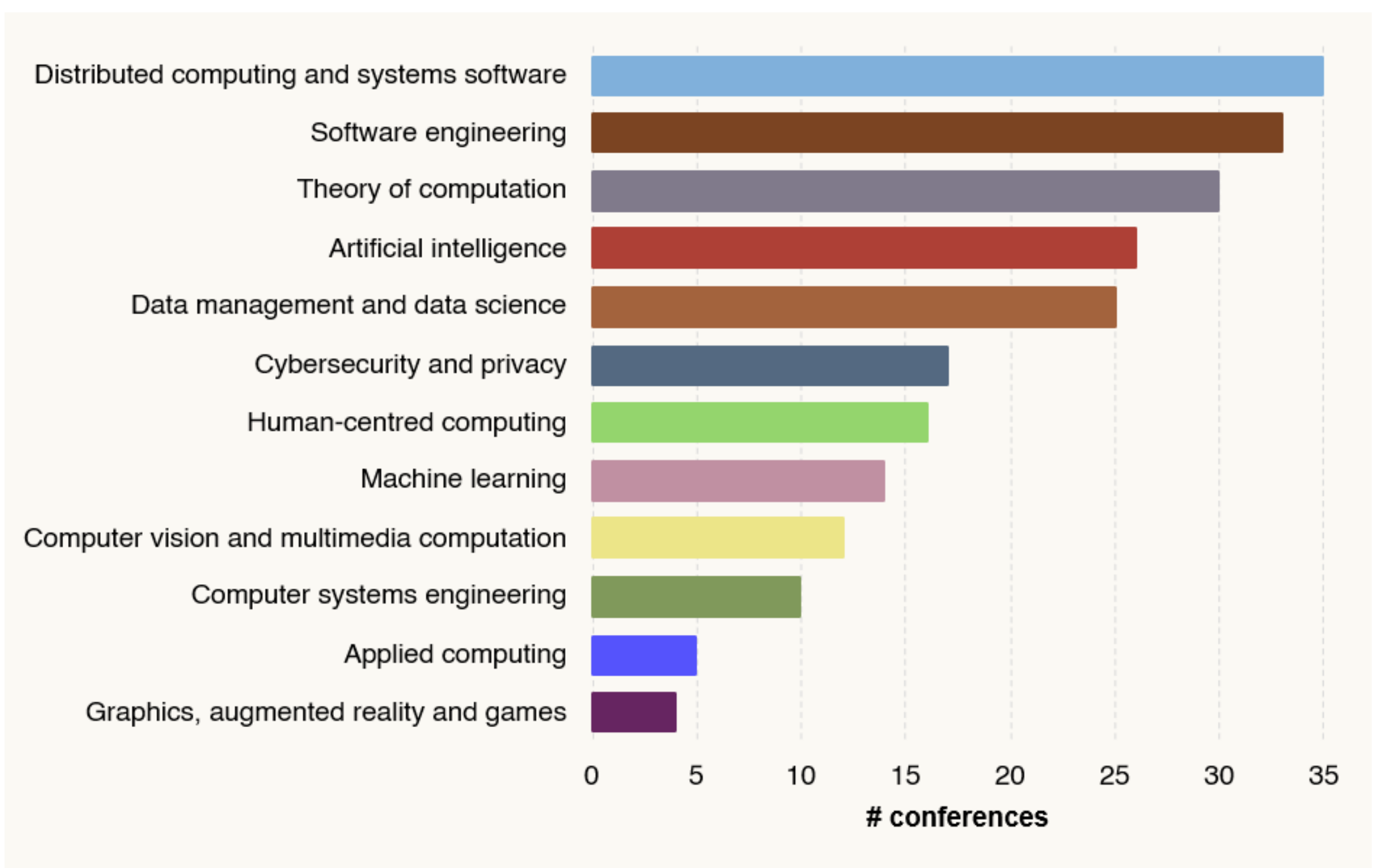

**Figure A1: Field distribution of the selected computer science conferences.**

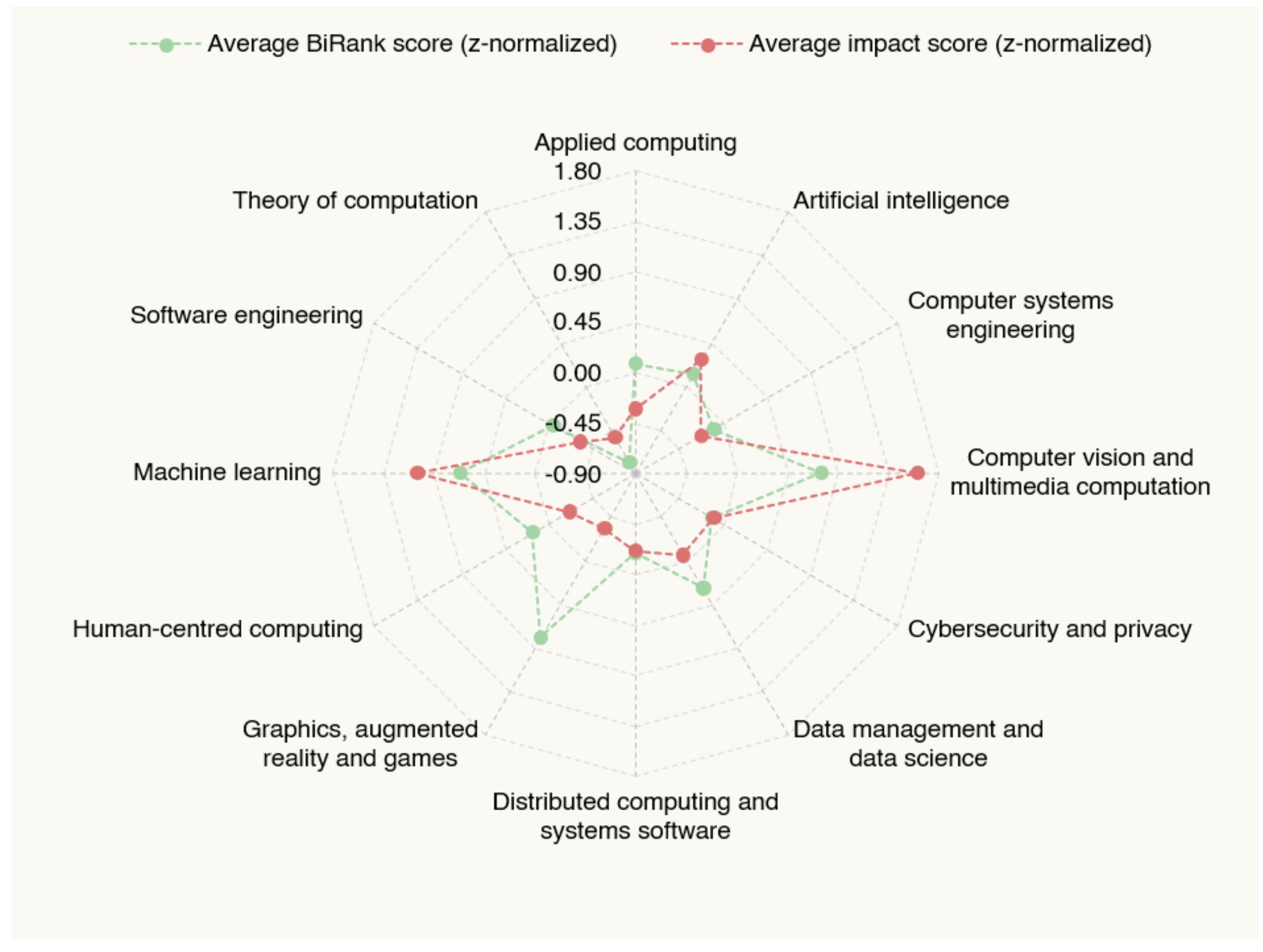

**Figure A2: Average BiRank scores versus impact scores across fields.**

## SUPPLEMENTARY INFORMATION

**Table S1: Selected computer science conferences.**

| No | Title | Acronym | Rank |
|---|---|---|---|
| 1 | National Conference of the American Association for Artificial Intelligence | AAAI | A* |
| 2 | International Joint Conference on Autonomous Agents and Multiagent Systems (previously the International Conference on Multiagent Systems, ICMAS, changed in 2000) | AAMAS | A* |
| 3 | Association for Computational Linguistics | ACL | A* |
| 4 | ACM Multimedia | ACMMM | A* |
| 5 | Automated Software Engineering Conference | ASE | A* |
| 6 | Architectural Support for Programming Languages and Operating Systems | ASPLOS | A* |
| 7 | Computer Aided Verification | CAV | A* |
| 8 | ACM Conference on Computer and Communications Security | CCS | A* |
| 9 | International Conference on Human Factors in Computing Systems | CHI | A* |
| 10 | Conference on Learning Theory | COLT | A* |
| 11 | Advances in Cryptology | CRYPTO | A* |
| 12 | IEEE Conference on Computer Vision and Pattern Recognition | CVPR | A* |
| 13 | ACM Conference on Economics and Computation | EC | A* |
| 14 | European Conference on Computer Vision | ECCV | A* |
| 15 | Empirical Methods in Natural Language Processing | EMNLP | A* |
| 16 | International Conference on the Theory and Application of Cryptographic Techniques | EuroCrypt | A* |
| 17 | IEEE Symposium on Foundations of Computer Science | FOCS | A* |

| No | Title | Acronym | Rank |
|---|---|---|---|
| 18 | ACM International Conference on the Foundations of Software Engineering (was ESEC/FSE, changed 2024; duplicate previously listed as ESEC, removed from DB) | FSE | A* |
| 19 | International Symposium on High Performance Computer Architecture | HPCA | A* |
| 20 | International Conference on Automated Planning and Scheduling | ICAPS | A* |
| 21 | IEEE International Conference on Computer Vision | ICCV | A* |
| 22 | International Conference on Data Engineering | ICDE | A* |
| 23 | IEEE International Conference on Data Mining | ICDM | A* |
| 24 | International Conference on Learning Representations | ICLR | A* |
| 25 | International Conference on Machine Learning | ICML | A* |
| 26 | IEEE International Conference on Robotics and Automation | ICRA | A* |
| 27 | International Conference on Software Engineering | ICSE | A* |
| 28 | International Joint Conference on Artificial Intelligence | IJCAI | A* |
| 29 | IEEE International Conference on Computer Communications | INFOCOM | A* |
| 30 | Information Processing in Sensor Networks | IPSN | A* |
| 31 | ACM International Symposium on Computer Architecture | ISCA | A* |
| 32 | IEEE/ACM International Symposium on Mixed and Augmented Reality | ISMAR | A* |
| 33 | ACM International Conference on Knowledge Discovery and Data Mining | KDD | A* |
| 34 | International Conference on the Principles of Knowledge Representation and Reasoning | KR | A* |
| 35 | IEEE Symposium on Logic in Computer Science | LICS | A* |
| 36 | ACM International Conference on Mobile Computing and Networking | MOBICOM | A* |
| 37 | Usenix Network and Distributed System Security Symposium | NDSS | A* |
| 38 | Advances in Neural Information Processing Systems (was NIPS) | NeurIPS | A* |

| No | Title | Acronym | Rank |
|---|---|---|---|
| 39 | Usenix Symposium on Operating Systems Design and Implementation | OSDI | A* |
| 40 | IEEE International Conference on Pervasive Computing and Communications | PERCOM | A* |
| 41 | ACM-SIGPLAN Conference on Programming Language Design and Implementation | PLDI | A* |
| 42 | ACM Symposium on Principles of Distributed Computing | PODC | A* |
| 43 | ACM SIGMOD-SIGACT-SIGART Conference on Principles of Database Systems | PODS | A* |
| 44 | ACM-SIGACT Symposium on Principles of Programming Languages | POPL | A* |
| 45 | Real Time Systems Symposium | RTSS | A* |
| 46 | ACM Conference on Embedded Networked Sensor Systems | SENSYS | A* |
| 47 | ACM Conference on Applications, Technologies, Architectures, and Protocols for Computer Communication | SIGCOMM | A* |
| 48 | ACM SIG International Conference on Computer Graphics and Interactive Techniques | SIGGRAPH | A* |
| 49 | ACM International Conference on Research and Development in Information Retrieval | SIGIR | A* |
| 50 | Measurement and Modeling of Computer Systems | SIGMETRICS | A* |
| 51 | ACM Special Interest Group on Management of Data Conference | SIGMOD | A* |
| 52 | ACM/SIAM Symposium on Discrete Algorithms | SODA | A* |
| 53 | ACM SIGOPS Symposium on Operating Systems Principles | SOSP | A* |
| 54 | IEEE Symposium on Security and Privacy | SP | A* |
| 55 | ACM Symposium on Theory of Computing | STOC | A* |
| 56 | ACM Symposium on User Interface Software and Technology | UIST | A* |
| 57 | Usenix Security Symposium | USENIX-Security | A* |
| 58 | International Conference on Very Large Databases | VLDB | A* |
| 59 | IEEE Conference on Virtual Reality and 3D User Interfaces | VR | A* |
| 60 | International World Wide Web Conference | WWW | A* |

| No | Title | Acronym | Rank |
|---|---|---|---|
| 61 | IEEE Real-Time and Embedded Technology and Applications Symposium | RTAS | A |
| 62 | IEEE International Conference on Software Analysis, Evolution and Reengineering | SANER | A |
| 63 | International Conference for High Performance Computing, Networking, Storage and Analysis (was Supercomputing Conference) | SC | A |
| 64 | SIAM International Conference on Data Mining | SDM | A |
| 65 | International Symposium on Software Engineering for Adaptive and Self-Managing Systems | SEAMS | A |
| 66 | ACM International Conference on Advances in Geographic Information Systems | SIGSPATIAL | A |
| 67 | Tools and Algorithms for Construction and Analysis of Systems | TACAS | A |
| 68 | Theoretical Aspects of Rationality and Knowledge | TARK | A |
| 69 | Theory of Cryptography Conference | TCC | A |
| 70 | Conference in Uncertainty in Artificial Intelligence | UAI | A |
| 71 | Usenix Annual Technical Conference | USENIX | A |
| 72 | IEEE Workshop on Applications of Computer Vision | WACV | A |
| 73 | International Conference on Logic Programming | ICLP | A |
| 74 | IEEE International Conference on Multimedia and Expo | ICME | A |
| 75 | IEEE International Conference on Program Comprehension (previously IWPC, changed in 2006) | ICPC | A |
| 76 | ACM International Conference on Supercomputing | ICS | A |
| 77 | International Conference on Software Architecture (was previously WICSA) | ICSA | A |
| 78 | IEEE International Conference on Software Maintenance and Evolution (prior to 2014 was ICSM, IEEE International Conference on Software Maintenance) | ICSME | A |
| 79 | International Conference on Service Oriented Computing | ICSOC | A |
| 80 | International Conference on Software Testing, Verification and Validation | ICST | A |

| No | Title | Acronym | Rank |
|---|---|---|---|
| 81 | IEEE Visualization | IEEE VIS | A |
| 82 | International Joint Conference on Automated Reasoning | IJCAR | A |
| 83 | Internet Measurement Conference | IMC | A |
| 84 | Interspeech (combined EuroSpeech and ICSLP in 2000) | Interspeech | A |
| 85 | IEEE International Parallel and Distributed Processing Symposium (was IPPS and SPDP) | IPDPS | A |
| 86 | International Symposium on Symbolic and Algebraic Computation | ISSAC | A |
| 87 | International Symposium on Software Reliability Engineering | ISSRE | A |
| 88 | International Symposium on Software Testing and Analysis | ISSTA | A |
| 89 | International Semantic Web Conference | ISWC | A |
| 90 | Innovations in Theoretical Computer Science | ITCS | A |
| 91 | International Conference on Intelligent User Interfaces | IUI | A |
| 92 | Logics in Artificial Intelligence, European Conference | JELIA | A |
| 93 | International Conference on Learning Analytics and Knowledge | LAK | A |
| 94 | International Symposium on Mathematical Foundations of Computer Science | MFCS | A |
| 95 | Medical Image Computing and Computer-Assisted Intervention | MICCAI | A |
| 96 | ACM Symposium on Mobile Ad Hoc Networking and Computing | MOBIHOC | A |
| 97 | International Conference on Model Driven Engineering Languages and Systems (Previously UML, changed in 2005) | MODELS | A |
| 98 | IEEE International Working Conference on Mining Software Repositories | MSR | A |
| 99 | ACM/IEEE International Conference on Modelling, Analysis and Simulation of Wireless and Mobile Systems | MSWIM | A |
| 100 | ACM Conference on Object Oriented Programming Systems Languages and Applications | OOPSLA | A |
| 101 | Privacy Enhancing Technologies Symposium (was International Workshop of Privacy Enhancing Technologies) | PETS | A |

| No | Title | Acronym | Rank |
|---|---|---|---|
| 102 | Parallel Problem Solving from Nature | PPSN | A |
| 103 | The International Symposium on Research in Attacks, Intrusions and Defenses (was International Symposium on Recent Advances in Intrusion Detection) | RAID | A |
| 104 | ACM International Conference on Recommender Systems | RecSys | A |
| 105 | European Conference on Machine Learning and Principles and Practice of Knowledge Discovery in Database (PKDD and ECML combined from 2008) | ECML PKDD | A |
| 106 | IEEE/IFIP International Conference on Dependable Systems and Networks | DSN | A |
| 107 | ACM SIGMOBILE International Conference on Mobile Systems, Applications and Services | Mobisys | A |
| 108 | IEEE International Conference on Web Services | ICWS | A |
| 109 | Conference on Interactive Theorem Proving (previously TPHOLs, changed in 2009) | ITP | A |
| 110 | International Conference on Concurrency Theory | CONCUR | A |
| 111 | ACM Virtual Reality Software and Technology | VRST | A |
| 112 | International Symposium on Theoretical Aspects of Computer Science | STACS | A |
| 113 | North American Association for Computational Linguistics | NAACL | A |
| 114 | British Machine Vision Conference | BMVC | A |
| 115 | International Symposium on Computational Geometry | SoCG | A |
| 116 | International Symposium on Algorithms and Computation | ISAAC | A |
| 117 | IEEE International Requirements Engineering Conference | RE | A |
| 118 | ACM/IEEE International Conference on Human-Robot Interaction | HRI | A |
| 119 | International Conference on Web and Social Media | ICWSM | A |
| 120 | IEEE/RSJ International Conference on Intelligent Robots and Systems | IROS | A |
| 121 | International Conference on Theory and Applications of Satisfiability Testing | SAT | A |

| No | Title | Acronym | Rank |
|---|---|---|---|
| 122 | European Association for Computational Linguistics | EACL | A |
| 123 | Conference on Integer Programming and Combinatorial Optimization | IPCO | A |
| 124 | International Computing Education Research Workshop | ICER | A |
| 125 | ACM Special Interest Group on Computer Science Education Conference | SIGCSE | A |
| 126 | ACM/IFIP/USENIX International Middleware Conference | Middleware | A |
| 127 | IEEE Conference on Computational Complexity | CCC | A |
| 128 | International Conference on Artificial Intelligence in Education | AIED | A |
| 129 | European Conference on Information Retrieval | ECIR | A |
| 130 | Annual Computer Security Applications Conference | ACSAC | A |
| 131 | European Symposium on Algorithms | ESA | A |
| 132 | ACM International Conference on Web Search and Data Mining | WSDM | A |
| 133 | International Conference on Artificial Intelligence and Statistics | AISTATS | A |
| 134 | Workshop on Algorithm Engineering and Experiments | ALENEX | A |
| 135 | International Workshop on Approximation Algorithms for Combinatorial Optimization Problems and International Conference on Randomization and Computation | APPROX/RANDOM | A |
| 136 | Asia Conference on Information, Computer and Communications Security | AsiaCCS | A |
| 137 | International Conference on the Theory and Application of Cryptology and Information Security | ASIACRYPT | A |
| 138 | International ACM SIGACCESS Conference on Computers and Accessibility | ASSETS | A |
| 139 | International Conference in Business Process Management | BPM | A |
| 140 | International Conference on Automated Deduction | CADE | A |
| 141 | International Conference on Advanced Information Systems Engineering | CaiSE | A |
| 142 | International Symposium on Code Generation and Optimization | CGO | A |

| No | Title | Acronym | Rank |
|---|---|---|---|
| 143 | Workshop on Cryptographic Hardware and Embedded Systems | CHES | A |
| 144 | Conference on Innovative Data Systems Research | CIDR | A |
| 145 | ACM International Conference on Information and Knowledge Management | CIKM | A |
| 146 | ACM International Conference on Emerging Networking Experiments and Technologies | CoNEXT | A |
| 147 | International Conference on Principles and Practice of Constraint Programming | CP | A |
| 148 | ACM Conference on Computer Supported Cooperative Work | CSCW | A |
| 149 | IEEE Computer Security Foundations Symposium (was CSFW) | CSF | A |
| 150 | Designing Interactive Systems | DIS | A |
| 151 | International Symposium on Distributed Computing (was WDAG) | DISC | A |
| 152 | International Conference on Evaluation and Assessment in Software Engineering | EASE | A |
| 153 | European Conference on Artificial Intelligence | ECAI | A |
| 154 | European Conference on Object-Oriented Programming | ECOOP | A |
| 155 | European Conference on Software Architecture | ECSA | A |
| 156 | Extending Database Technology | EDBT | A |
| 157 | International Conference on Conceptual Modelling | ER | A |
| 158 | International Symposium on Empirical Software Engineering and Measurement | ESEM | A |
| 159 | European Symposium on Programming | ESOP | A |
| 160 | European Symposium On Research In Computer Security | ESORICS | A |
| 161 | IEEE European Symposium on Security and Privacy | EuroS&P | A |
| 162 | Eurosys Conference | EuroSys | A |
| 163 | Conference on File and Storage Technologies | FAST | A |
| 164 | Financial Cryptography and Data Security Conference | FC | A |

| No | Title | Acronym | Rank |
|---|---|---|---|
| 165 | International Symposium on Formal Methods (was Formal Methods Europe FME) | FM | A |
| 166 | Foundations of Genetic Algorithms | FOGA | A |
| 167 | Foundations of Software Science and Computational Structures | FOSSACS | A |
| 168 | Graph Drawing | GD | A |
| 169 | Genetic and Evolutionary Computations | GECCO | A |
| 170 | USENIX Workshop on Hot Topics in Operating Systems | HotOS | A |
| 171 | ACM International Symposium on High Performance Distributed Computing | HPDC | A |
| 172 | International Colloquium on Automata Languages and Programming | ICALP | A |
| 173 | IEEE/ACM International Conference on Computer-Aided Design | ICCAD | A |
| 174 | IEEE International Conference on Document Analysis and Recognition | ICDAR | A |
| 175 | International Conference on Distributed Computing Systems | ICDCS | A |
| 176 | International Conference on Database Theory | ICDT | A |
| 177 | International Conference on Functional Programming | ICFP | A |

Source: CORE2023 from the ICORE Conference Portal.